\documentclass[twocolumn,superscriptaddress,prl,showpacs,preprintnumbers,amsmath,amssymb,floats]{revtex4-1}
\usepackage{graphicx}
\usepackage{amsmath}
\usepackage{epstopdf}
\usepackage{amsbsy}
\usepackage{color}
\usepackage{lipsum}
\usepackage{dcolumn}
\usepackage{bm}
\usepackage{dsfont}

\newcommand{\kkk}{\mathbf{k}}

\newcommand{\RRR}{\mathbf{R}}
\newcommand{\rrr}{\mathbf{r}}
\newcommand{\qqq}{\mathbf{q}}

\newcommand{\ppp}{\mathbf{p}}
\newcommand{\mbf}[1]{\mathbf{#1}}

\begin{document}
\title{Magnetic fluctuations in pair density wave superconductors }

\author{Morten H. Christensen}

\affiliation{Niels Bohr Institute, University of Copenhagen, Juliane Maries Vej 30, DK-2100 Copenhagen,
Denmark}

\author{Henrik Jacobsen}

\affiliation{Niels Bohr Institute, University of Copenhagen, Juliane Maries Vej 30, DK-2100 Copenhagen,
Denmark}

\author{Thomas A. Maier}

\affiliation{Computer Science and Mathematics Division and Center for Nanophase
Materials Sciences, Oak Ridge National Laboratory, Oak Ridge, Tennessee 37831, USA}

\author{Brian M. Andersen}

\affiliation{Niels Bohr Institute, University of Copenhagen, Juliane Maries Vej 30, DK-2100 Copenhagen,
Denmark}

\date{\today}

\begin{abstract}

Pair density wave superconductivity constitutes a novel electronic condensate
proposed to be realized in certain unconventional superconductors.
Establishing its potential existence is important for our fundamental
understanding of superconductivity in correlated materials. Here we compute
the dynamical magnetic susceptibility in the presence of a pair density wave
ordered state, and study its fingerprints on the spin-wave
spectrum including the neutron resonance. In contrast to the standard case of
$d$-wave superconductivity, we show that the pair density wave phase exhibits
neither a spin-gap nor a magnetic resonance peak, in agreement with a recent neutron scattering experiment on
underdoped La$_{1.905}$Ba$_{0.095}$CuO$_4$ [Z. Xu \textit{et al.}, Phys.
Rev. Lett. {\bf 113}, 177002 (2014)].

\end{abstract}

\maketitle

Superconductivity can have significant effects on the structure of the
spin fluctuations. This includes, for example, the opening of a spin-gap at low energies and
the appearance of a magnetic neutron resonance when the gap exhibits sign changes along the Fermi surface as in cuprates and iron-based materials~\cite{scalapino12}. Similarly, the structure of the magnetic fluctuations can
have important consequences for the superconducting state, even possibly its
mere existence~\cite{berk66}. Thus, spin fluctuations and unconventional
superconductivity are intimately linked, and the question of exactly how they
are connected and what this tells us about the pairing mechanism~\cite{scalapino12} remains a challenging and relevant problem in the
field of high-temperature superconductivity.

The pseudogap regime of the underdoped cuprates is highly susceptible to spin
and charge order. Unidirectionally (striped) modulated spin and charge order
was first discovered near a hole doping of $x = 1/8$ in
La$_{1.6-x}$Nd$_{0.4}$Sr$_x$CuO$_4$ \cite{tranquada95}, and subsequently in
other cuprates also exhibiting low-temperature tetragonal crystal structure,
including La$_{2-x}$Ba$_x$CuO$_4$
\cite{klauss00,fujita04,fink09,hucker11}. However, stripe correlations appear to
be present in many other cuprates, and the universal hour-glass
spin excitation spectrum observed in inelastic neutron scattering
experiments has been explained within stripe models
\cite{kivelson03,cupratereview1,cupratereview2}. On the other hand,
calculations based on purely itinerant models that include $d$-wave
superconductivity but no static stripe order also find a neutron resonance with an
hour-glass dispersion~\cite{Eschrig2006}.  At present, a detailed
quantitative description of the spin dynamics of the cuprates, and its evolution from
antiferromagnetic spin waves in the parent compounds to itinerant paramagnons
with a clear spin-gap and a neutron resonance in the overdoped regime, remains
an unsettled problem. Hence it is important to study the intermediate doping
regime where prominent stripe correlations coexist with superconducting order.

\begin{figure}[t]
\centering
\includegraphics[width=\columnwidth]{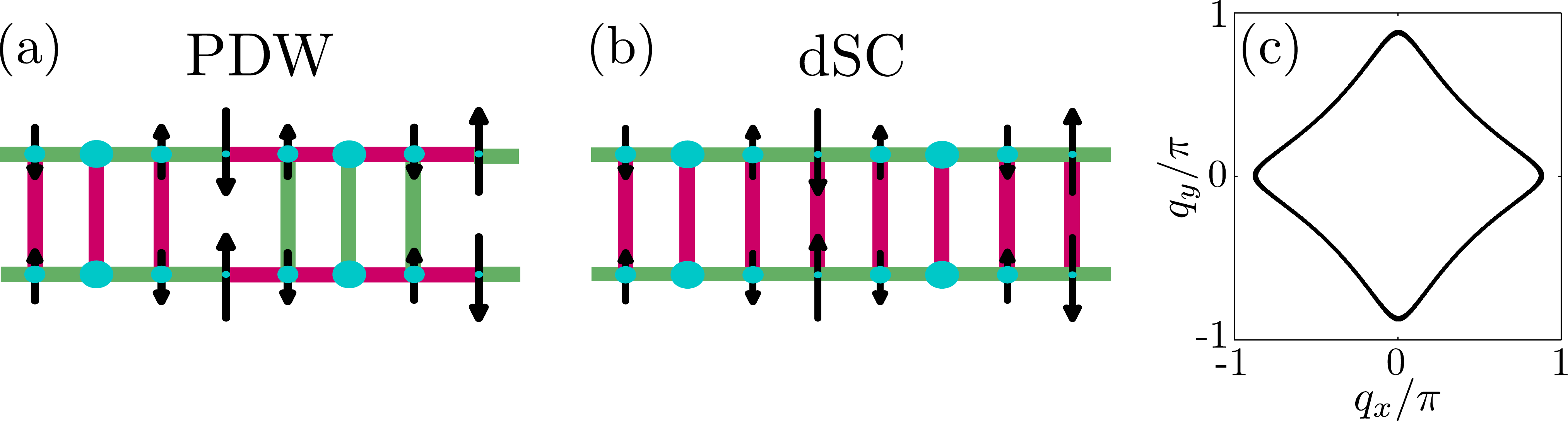}
\caption{\label{fig:unit_cells} (Color online)  (a)-(b) Illustration of the unit cells with charge and spin order, and either PDW (a) or dSC (b) 
superconducting order. The arrows denote the magnetization, the diameter of the circles
the hole density, and the colors on the bonds indicate the sign of the
superconducting order parameter, green is positive and magenta is negative. (c) Normal state Fermi surface with a doping of $12.5\%$.}
\end{figure}

An experimental study of the transport properties of striped La$_{1.875}$Ba$_
{0.125}$CuO$_4$ \cite{li07} reported 2D superconductivity coexisting with
stripe order at temperatures above the 3D superconducting transition
temperature. This was taken as evidence for an anti-phase ordering of the
superconducting order parameter between the CuO$_2$ layers, which suppresses the
inter-layer Josephson coupling required for 3D superconductivity. The
existence of pair density wave (PDW) order, in which striped charge,
magnetic and superconducting orders are intertwined with unusual sign changes of the superconducting phase~\cite{agterberg08,barush08}, has been proposed as a
possible explanation of these findings \cite{Berg07,berg09}. In the PDW state, the
superconducting order parameter has a finite Cooper pair momentum with periodicity equal to that of the magnetic stripe order as illustrated in Fig.~\ref{fig:unit_cells}(a)
\cite{Fradkin15}. This is to be contrasted with a more ordinary modulated $d$-wave superconductor (dSC), in which the superconducting order parameter is in-phase across the
stripe domains, and modulated in amplitude with the same periodicity as the charge stripes, i.e. half the wave length of the PDW state as shown in Fig.~\ref{fig:unit_cells}(b).

The possibility of a PDW state was investigated within microscopic models, and numerical studies of the $t$-$J$ model generally find that this state is energetically competitive
with other more ordinary modulated superconducting states~\cite{Himeda02,Raczkowski07,Yang09,Corboze14}. Similar conclusions were reached within an extended version of BCS theory above a critical pairing strength~\cite{loder10}, while subsequent Hartree-Fock studies focused on the single-particle electronic properties of phases of combined PDW order and AFM stripes~\cite{loder11,loder11njp}. More recently, finite momentum superconducting PDW order has resurfaced in theoretical studies of the charge density wave (CDW) order detected in underdoped cuprates~\cite{lee14,Wang14,Pepin14}. The existence of an entangled CDW/PDW phase was found and analyzed both in the context of an emergent SU(2) symmetry of the fermionic hot-spot model~\cite{Pepin14,Freire15}, and in the spin-fermion model close to the onset of antiferromagnetism~\cite{Wang15,Wang15_2}.

Experimentally, a recent neutron scattering study of the 
low-energy spin response in stripe ordered La$_{1.905}$Ba$_{0.095}$CuO$_4$ \cite
{xu14} found a number of remarkable results that were taken as evidence for a
PDW state: (1) gapless spin excitations coexisting with superconductivity,
and (2) the absence of a neutron resonance in the superconducting state. These
results are highly unusual since both a spin-gap and a neutron resonance are
expected in unconventional superconductors like the cuprates~\cite{scalapino12}. 

Motivated by the experimental findings of Ref.~\onlinecite{xu14}, we perform a theoretical study of the
fingerprints of a putative PDW state on the inelastic neutron scattering
spectrum. We focus on the consequences of the PDW state rather than its microscopic origin. We find that the PDW state in the \textit{absence} of magnetic and
charge order exhibits neither a spin-gap nor a neutron resonance, contrary to the standard dSC phase. For the
state where PDW superconductivity coexists with striped magnetic order, we find
qualitatively similar results. In particular, the neutron scattering spectrum
in this coexistence phase is almost identical to that of the normal state. In the standard dSC phase, on the other hand, we show that the neutron resonance is robust to coexisting stripe order. These findings support a scenario where the absence of a spin-gap and a
magnetic resonance in underdoped La$_ {1.905}$Ba$_{0.095}$CuO$_4$\cite{xu14}
is explained by the existence of a PDW condensate.

The stripe phase coexisting with superconductivity is studied within a phenomenological mean-field one-band Hubbard model
\begin{eqnarray}\label{eq:mf_hamilton}
	\mathcal{H}^{\text{MF}} &=& -\sum_{ij\sigma}\left(t_{ij} + \mu \delta_{ij}
	\right)c^{\dagger}_{i\sigma}c_{j\sigma} + U\sum_{i\sigma}\langle
	n_{i\bar{\sigma}} \rangle n_{i\sigma} \nonumber \\ && - \sum_{\langle ij
	\rangle} \left[ \Delta_{ji} c^{\dagger}_{i\uparrow}c^{\dagger}_{j\downarrow} +
	\mbox{H.c.}\right]\,,
\end{eqnarray}
with $U>0$. For the hopping integrals $t_{ij}$, we include NN $t=1$ (setting the unit of energy) and NNN couplings $t'=-0.3$. The details of the bandstructure are not important for the results discussed below. The associated Fermi surface of the tight-binding model is depicted in Fig.~\ref{fig:unit_cells}(c). The Hamiltonian (\ref{eq:mf_hamilton}) and its generalizations have been used previously to study the stripe phase of the cuprates~\cite{zhu02,Chen,andersen05,Andersen10}, including the electronic properties of the PDW phase~\cite{loder11,loder11njp}, but an analysis of the spin response in the PDW phase has not previously been addressed theoretically.

Here we use an $8 \times 2$ supercell to study the effects of stripe and PDW order on the magnetic excitation spectrum. The periodicity of the magnetic (charge) stripe order is therefore restricted to 8 (4) lattice sites along $\hat{x}$ and 2 (1) sites along $\hat{y}$. This restriction limits the possible solutions and a selfconsistent iterative procedure in general only obtains a saddle point in the free energy landscape. The actual minimum is often located at a different periodicity which is inaccessible due to the restriction to $8\times 2$ periodic unit cells. In such cases the Goldstone modes either remain gapped or the spin-wave branches cross zero energy before reaching the ordering vector~\cite{dallapiazza15}. To study the spin response in the presence of $8 \times 2$ periodic stripes we therefore adopt an alternative approach: we impose a density modulation $\langle n_{i\sigma} \rangle$, corresponding to site-ordered magnetic and charge stripes, and a superconducting order parameter, $|\Delta_{ij}|=0.05$, corresponding to either dSC or PDW order, as shown in Fig.~\ref{fig:unit_cells}(a)-(b). For each chosen configuration, we subsequently adjust the bare interaction $U$ such that Goldstone's theorem is satisfied, i.e. such that the denominator of the real part of the RPA susceptibility exhibits a zero eigenvalue at $q_x=\pi\pm \frac{\pi}{4}$ (see the Supplementary Material (SM)~\cite{supplementary} for further details). This procedure guarantees a stable energy minimum in the energy landscape of $8 \times 2$ periodic stripes, and has the benefit of allowing us to study PDW-, dSC-, and non-superconducting solutions within the same region of parameter space and the same assumed density modulations. This allows us to single out the effects of just the PDW order on the spin susceptibility.

We apply a supercell formalism, where the total $N \times N$ (here $N=96$) system consists of supercells of size $8 \times 2$.
The dynamical spin susceptibility $\chi^{+-}
(\qqq,\omega)=\chi^{+-}(\qqq,i\omega_n\rightarrow\omega+i\delta)$ that
determines the neutron scattering intensity may be obtained from
\begin{eqnarray}
	\chi^{+-}(\qqq,i\omega_n)&=&\sum_{\rrr_i,\rrr_j} e^{-i\qqq
	(\rrr_i-\rrr_j)}\chi^{+-}(\rrr_i,\rrr_j,i\omega_n),
\end{eqnarray}
which contains terms originating from both the intra- and inter-supercell structure. Here $\mbf{r}_i=\mbf{R}_i + i$ where $\mbf{R}_i$ yields the supercell containing site $\mbf{r}_i$ and $i$ gives the site of $\mbf{r}_i$ in that supercell. The site dependent susceptibility is obtained from
\begin{eqnarray}
	\chi^{+-}(\rrr_i,\rrr_j,i\omega_n)=\int_0^\beta d\tau e^{i\omega_n\tau}
	\left\langle
	S^{+}(\rrr_{i},\tau)S^{-}(\rrr_{j},0) \right\rangle \,,\nonumber
\end{eqnarray} 
where $S^{+}(\rrr_{i},\tau) = c^\dagger_{\mbf{r}_i \uparrow}(\tau)c^{\phantom\dagger}_
{\mbf{r}_i \downarrow}(\tau)$ is the spin raising operator at position $\rrr_{i}$ at
(imaginary) time $\tau$ and $S^ {-} (\rrr_ {i},\tau)$ the corresponding spin lowering
operator. The bare susceptibility takes the standard form, consisting of
contributions from both normal and anomalous Green functions as detailed
in the (SM)~\cite{supplementary}. At the RPA level, the site-dependent
susceptibility is given by
\begin{eqnarray}
	\chi^{+-}(\rrr_i,\rrr_j,\omega) &=& \chi^{+-}_0(\rrr_i,\rrr_j,\omega) \\
	&+& U\sum_{\rrr_l}\chi^{+-}_0(\rrr_i,\rrr_l,\omega)\chi^{+-}
	(\rrr_l,\rrr_j,\omega)\,,\nonumber
\end{eqnarray}
where $\chi^{+-}_0(\rrr_i,\rrr_j,\omega)$ is the bare susceptibility
calculated with respect to the mean-field Hamiltonian in
Eq.~(\ref{eq:mf_hamilton}).

\begin{figure}[b]
\centering
\includegraphics[width=\columnwidth]{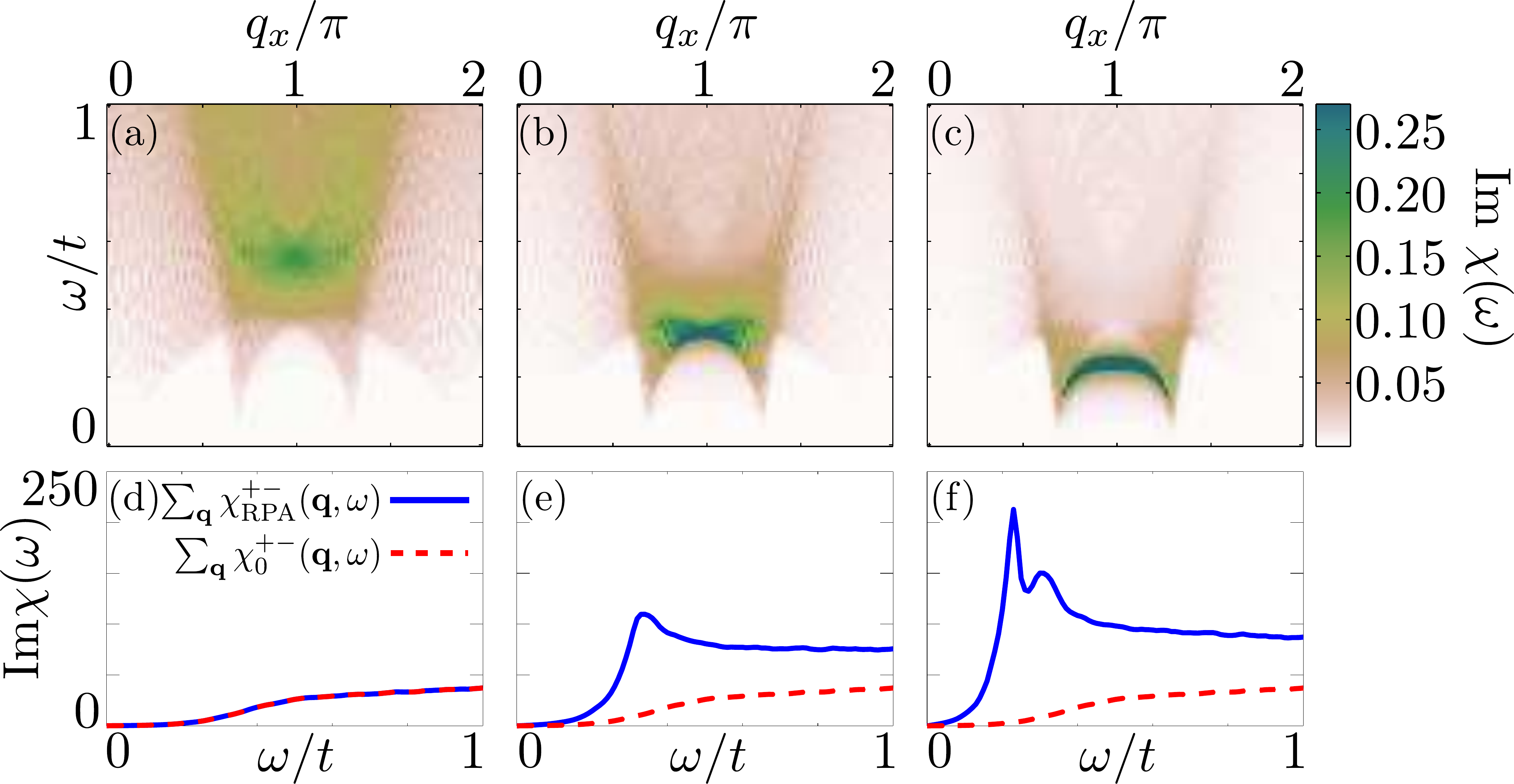}
\caption{\label{fig:dSC} (Color online) (a)-(c) Imaginary part of the spin susceptibility
$\mathrm {Im}\chi^{+-}(\mbf{q},\omega)$ with only dSC order. Here we have set $q_y=\pi$ and plot
$\mathrm {Im}\chi^{+-}(q_x,\pi,\omega)$ versus $q_x/\pi$ and $\omega/t$ with (a)
$U=0$, (b) $U/t=1.8$, and (c) $U/t=2.2$. For clarity the intensity of the first two cases has been rescaled. (d)-(e) show the imaginary part of the $\mbf{q}$-integrated bare (red dashed) and RPA (blue solid lines) susceptibilities.}
\end{figure}

In order to disentangle the effects of superconductivity on
the spin response from those of the striped magnetic order, we
start by considering systems with either only dSC or only PDW order, i.e. without coexisting 
charge and magnetic order. In Fig.~\ref{fig:dSC} we show the imaginary part of
the RPA susceptibility for the dSC phase at $q_y = \pi$. As evident from the results for the bare susceptibility $\chi_0$ (red dashed curve) in Fig.~\ref{fig:dSC}(c)-(d), one clearly
sees the opening of a spin-gap below $2\Delta$ (at $U=0$). At finite $U$ a
resonance peak, which shifts to lower energies as $U$ increases, appears at energies slightly below the bare spin-gap as seen more clearly from Fig.~\ref{fig:dSC}(e)-(f), as expected for a superconducting gap that changes sign under translation of ${\mathbf{Q}}=(\pi,\pi)$~\cite{scalapino12,Eschrig2006}.

\begin{figure}[t]
\centering
\includegraphics[width=\columnwidth]{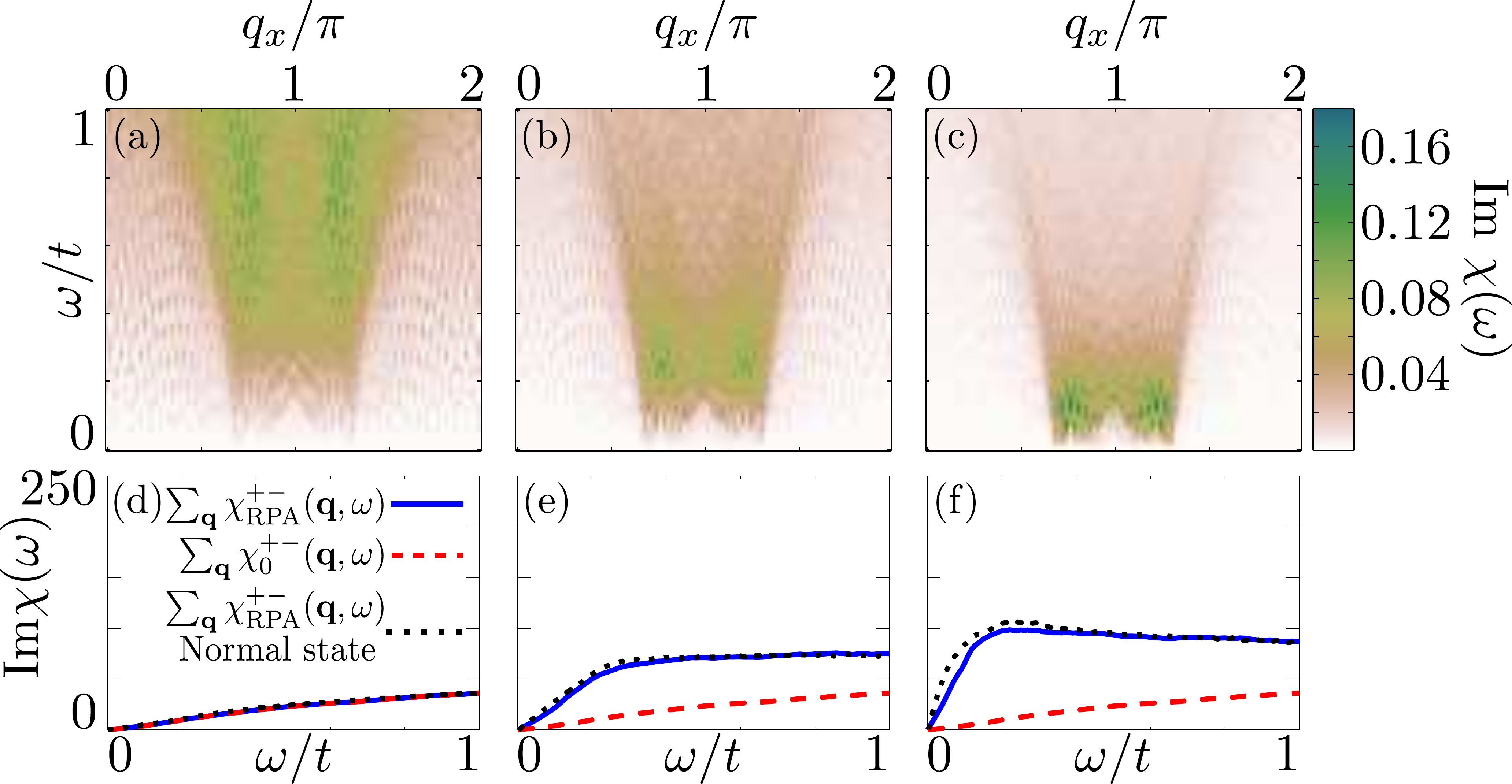}
\caption{\label{fig:PDW} (Color online) (a)-(c) Imaginary part of the spin susceptibility
$\mathrm {Im}\chi^{+-}(\mbf{q},\omega)$ versus $q_x/\pi$ and $\omega/t$ for a system with only PDW order for the same values of $U$ as in Fig.~\ref{fig:dSC}. (d)-(f) Imaginary part of the $\mbf{q}$-integrated susceptibilities corresponding to (a)-(c). The black dashed lines show the integrated RPA susceptibility in the normal state.}
\end{figure}

The corresponding results for the case with only PDW order are plotted in
Fig.~\ref{fig:PDW}, and seen to be in stark contrast to the phase with only dSC order (Fig.~\ref{fig:dSC}). In the
PDW phase, although the system is superconducting, a spin-gap is clearly
absent. Without a spin-gap, quasiparticle damping is not suppressed which
further implies that a magnetic resonance should be absent, consistent
with the RPA results displayed in Fig.~\ref{fig:PDW}. As seen, the spectral weight is rather structureless and
distributed over a wider range in both frequency and momentum. A comparison of the PDW phase with the normal (non-ordered) case, shown in Fig.~\ref{fig:PDW}(d-f) by the dotted black lines, reveals that the spin response of the normal state and the PDW state are in fact remarkably similar. 

\begin{figure}[b]
\centering
\includegraphics[width=0.5\textwidth]{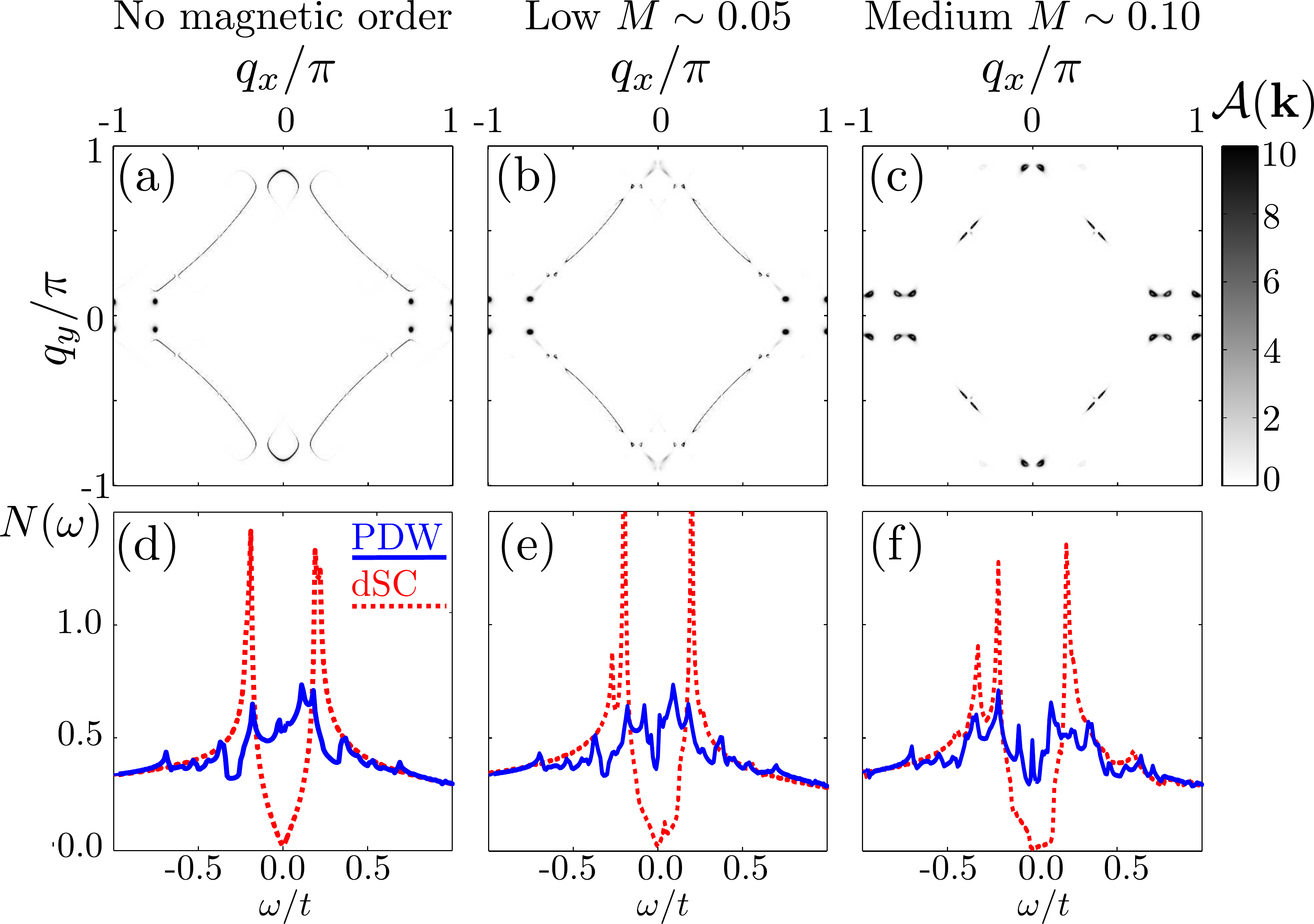}
\caption{\label{fig:spectral_func_and_dos} (Color online) (a)-(c) The spectral function $\mathcal{A}(\mbf{k},\omega=0)$ for a system with PDW order and increasing magnitude of the site-averaged magnetic moment. (d)-(f) Comparison of the DOS for the PDW and dSC phases for the same parameters corresponding to panels (a)-(c).}
\end{figure}

One may understand the absence of a spin-gap in the PDW state from the zero frequency single-particle spectral weight $\mathcal{A}(\mbf{k},\omega=0)$ and the associated density of states (DOS) displayed in Fig.~\ref{fig:spectral_func_and_dos}(a) and \ref{fig:spectral_func_and_dos}(d). The dSC phase (not shown) exhibits the usual gap structure, with gap nodes along the $|k_x|=|k_y|$ lines in $\mathcal{A}(\mbf{k})$. In contrast, the PDW phase exhibits states on large parts of the
Fermi surface [Fig.~\ref{fig:spectral_func_and_dos}(a)] and the DOS clearly does not exhibit a suppression of states near the Fermi level~\cite{loder11njp}. The low-energy states
in the PDW state are caused by the mismatch of the real-space
pairing bonds seen in Fig.~\ref{fig:unit_cells}(a), which are known to produce low-energy Andreev-like
zero-energy states~\cite{Andersen06}.

We now turn to the full coexistence phase with $8 \times 2$-periodic magnetic and charge stripes as well as PDW or dSC orders present.
Combining superconductivity with magnetic and charge order leads to a
reconstruction of the Fermi surface, evidenced in Fig.~\ref{fig:spectral_func_and_dos}(b,c). The effect of a finite (weak) magnetization on the DOS is
relatively minor, as seen in Fig. \ref{fig:spectral_func_and_dos}(e)-(f). The
system with a PDW does not exhibit a full gap, even at $\omega=0$, while the
gap present in the dSC case is only altered quantitatively by the addition of
magnetism. Similar conclusions hold for the spectral function: the PDW state still
exhibits states on large parts of the Fermi surface, while only states along
the nodal lines are present in the dSC phase.

\begin{figure}
\centering
\includegraphics[width=0.5\textwidth]{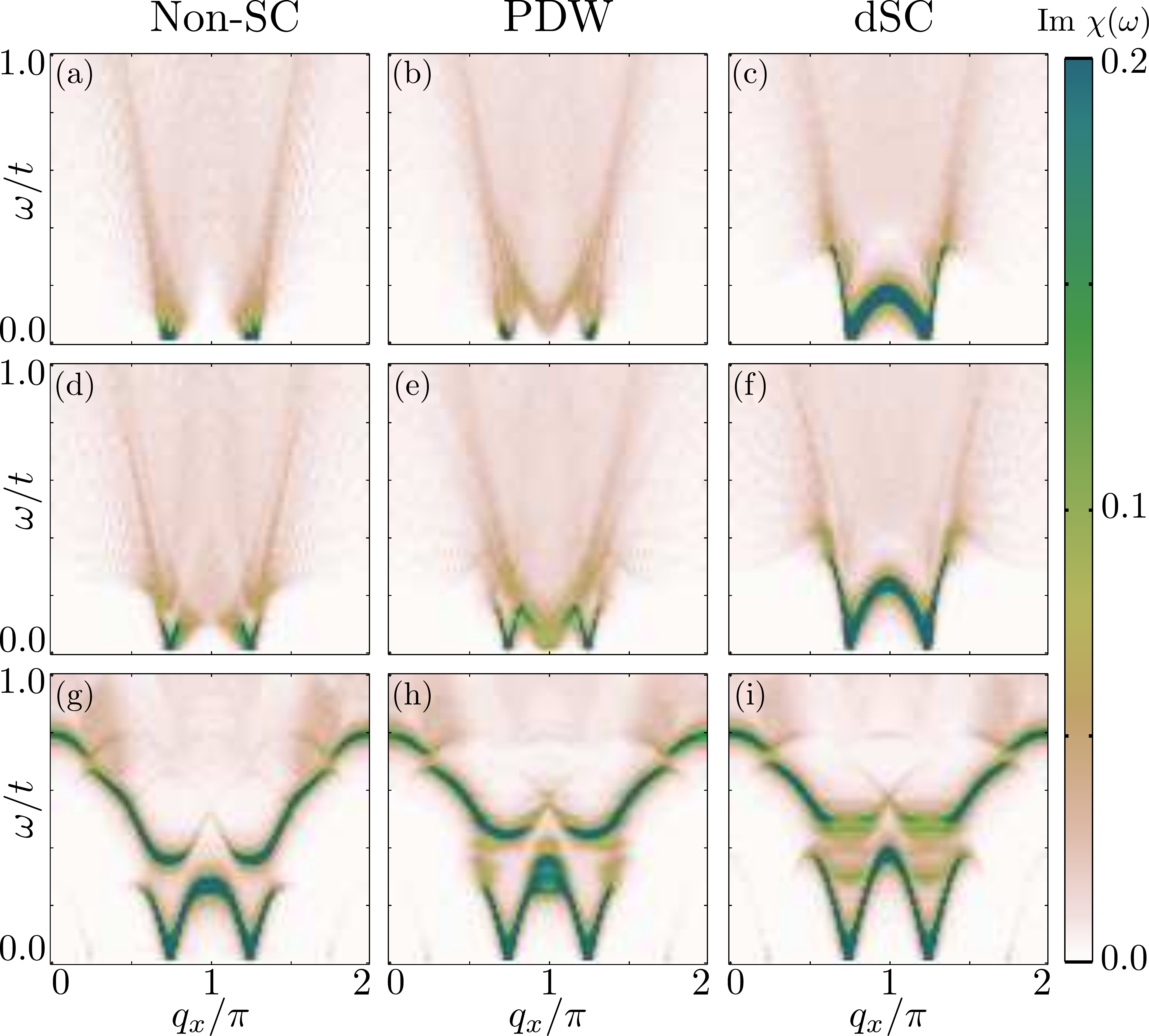}
\caption{\label{fig:med_M_spin_waves} (Color online) Imaginary part of the spin susceptibility
Im$\chi^{+-}(\qqq,\omega)$ in the presence of stripe charge and magnetic order without superconductivity (left column), and with superconductivity (middle and rightmost columns), for increasing magnetic order; (a)-(c) show the case where the site-averaged magnetic moment $M \sim 0.05$, (d)-(f) corresponds to $M \sim 0.1$, and (h)-(i) has $M \sim 0.4$.}
\end{figure}

Proceeding to study the spin-wave spectrum of the coexistence phase, we first
note that the presence of Goldstone modes necessarily excludes the opening of a spin-gap.
This is clearly seen in Fig. \ref{fig:med_M_spin_waves}, where we show the
imaginary part of the susceptibility $\chi^{+-}(\qqq,\omega)$ versus $q_x$ with
$q_y=\pi$  for an increasing magnitude of the site-averaged magnetic moment.
The Goldstone modes are seen by the high intensity peaks at $\omega=0$ for $q_x =
\pi \pm \pi/4$ for all the cases shown. In Fig.~\ref{fig:med_M_spin_waves}, panels (a)-(c) corresponds to a site-averaged magnetic moment of $M \sim 0.05$, panels (d)-(f) has $M \sim 0.1$~\cite{footnote}, and panels (g)-(i) has $M \sim 0.4$ (see the SM~\cite{supplementary} for the exact order parameters used). In the dSC phase, there is still a
resonance indicated by the region of high intensity at $q_x=\pi$ visible as the region of high intensity bridging the two spin wave branches as seen most clearly in Figs.~\ref{fig:med_M_spin_waves}(c,f). This is in stark contrast to the PDW case [Figs.~\ref{fig:med_M_spin_waves}(b,e)] where this coherent excitation is completely washed out, similar to the case shown in Fig.~\ref{fig:PDW} without charge and spin order. For larger magnetic moments the magnetic excitations approach the standard spin-wave branches of the stripe phase~\cite{kruger03,seibold06,yao06}, but a significantly broadened dispersion at the resonance point ($q_x=\pi$) is seen to remain present in the PDW phase compared to the dSC phase, as seen by comparison of Fig.~\ref{fig:med_M_spin_waves}(h)-(i).

\begin{figure}
\centering
\includegraphics[width=0.45\textwidth]{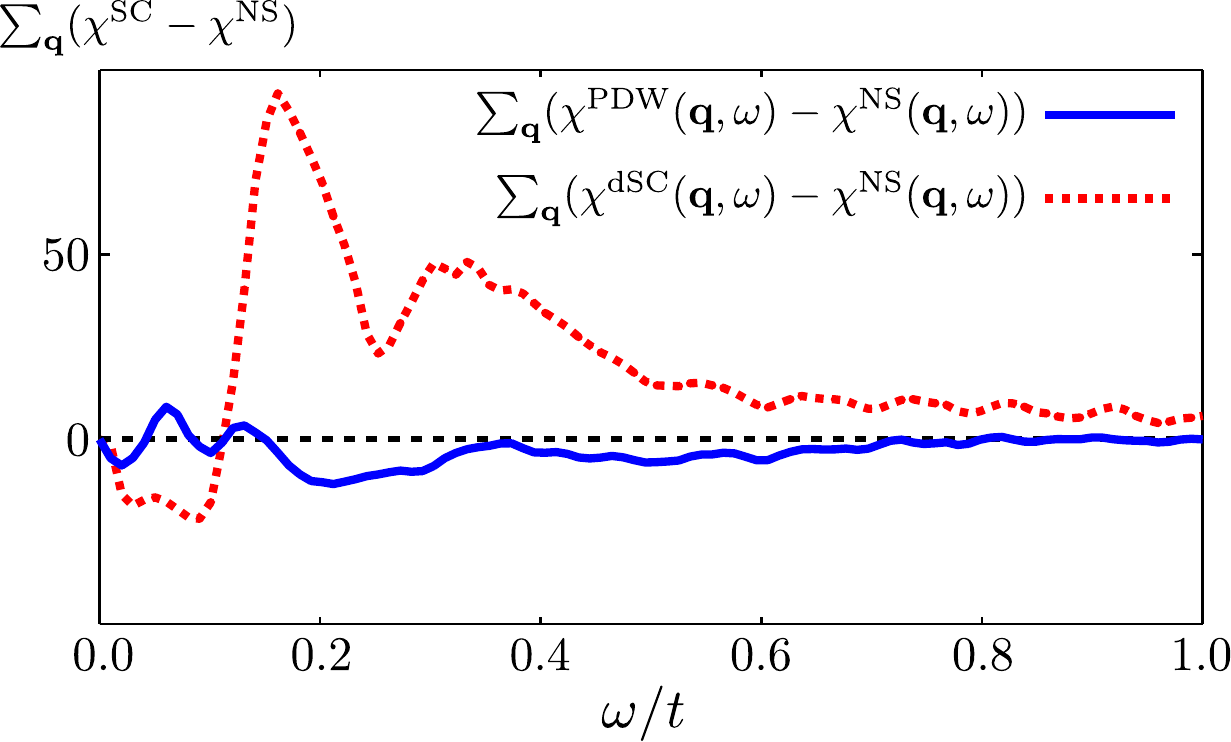}
\caption{\label{fig:resonance_difference} (Color online) Comparison of the integrated neutron
resonance for the PDW and dSC cases. The blue curve shows the difference
between the PDW and the non-superconducting case, while the red curve shows
the difference between the dSC case and the non-superconducting case. The dSC
shows clear signs of a resonance, which is absent for the PDW case.}
\end{figure}

To illustrate this more clearly, we show in Fig.~\ref{fig:resonance_difference} the
difference in the $\qqq$-integrated spin susceptibility between the
superconducting and normal state for both the PDW (solid blue) and dSC orders (dotted red).
These results are for the case where the site-averaged magnetic moment $M \sim 0.1$ [cases (d)-(f) in Fig.~\ref{fig:med_M_spin_waves}]. As seen, the dSC phase exhibits a clear resonance around $w/t \sim 0.18$, while the PDW phase is structureless. In fact, the PDW case has an almost
identical spin response to the normal state, a result that is in good agreement
with the experimental data measured on LBCO at $x=0.095$ by Xu {\it et al} \cite{xu14}.

To summarize, we have studied the distinct signatures of a PDW state
with intertwined striped spin, charge, and anti-phase superconducting bond order on the dynamic spin susceptibility. We find that in the PDW state both a
spin-gap and a neutron resonance are absent, in contrast to the coexistence phase with standard
in-phase $d$-wave superconductivity where the neutron resonance is preserved. This absence of the usual
fingerprint of a sign-changing superconducting gap in the PDW state can be traced
back to its gapless single-particle excitation spectrum. These results are in agreement with recent neutron scattering results on $x=0.095$ LBCO~\cite{xu14}, where neither a spin-gap nor a resonance were observed below the superconducting critical temperature.

\begin{acknowledgments}
The authors are grateful to W. A. Atkinson, P. J. Hirschfeld, A. P. Kampf, A. Kreisel, D. D. Scherer, F. Loder, N. Perkins, I. Rousochatzakis, A. T. R\o mer, J. Tranquada, and G. Yu for useful discussions. MHC and
BMA acknowledge financial support from a Lundbeckfond fellowship (grant
A9318). TAM acknowledges support from the Center of Nanophase Materials
Sciences, which is a DOE Office of Science User Facility.
\end{acknowledgments}


\newpage

\begin{widetext}

\begin{center}
	{\large\textbf{Supplementary Material to \\ ``Magnetic fluctuations in pair-density wave superconductors''}} \vspace{4mm} \newline
	\vspace{2mm}
	Morten H. Christensen$^{1}$, Henrik Jacobsen$^{1}$, Thomas A. Maier$^{2}$, and Brian M. Andersen$^{1}$ \newline
	\textit{{}$^{1}$ Niels Bohr Institute, University of Copenhagen, \newline Juliane Maries Vej 30, DK-2100 Copenhagen, Denmark}\newline
	\textit{{}$^{2}$ Computer Science and Mathematics Division and Center for Nanophase
Materials Sciences, \newline Oak Ridge National Laboratory, Oak Ridge, Tennessee 37831, USA}
\end{center}

\section{Transverse magnetic susceptibility with supercells}

Here we provide the details of the derivation of the transverse susceptibility in real-space on a superconducting ground state. The bare transverse susceptibility is
\begin{eqnarray}
	\chi^{+-}_{0}(\mbf{r}_{i},\mbf{r}_{j},\tau) = \left\langle T_{\tau} S^{+}(\mbf{r}_i,\tau)S^{-}(\mbf{r}_j,0)  \right\rangle_0\,,
\end{eqnarray}
where $S^{+}(\mbf{r}_i,\tau)$ [$S^{-}(\mbf{r}_i,\tau)$] is the spin creation [annihilation] operator at position $\mbf{r}_i$ at (imaginary) time $\tau$,
\begin{eqnarray}
	S^{+}(\mbf{r}_i,\tau)=c^{\dagger}_{\mbf{r}_i \uparrow}(\tau)c_{\mbf{r}_i \downarrow}(\tau)\,.
\end{eqnarray}
Applying the Bogoliubov-de Gennes (BdG) transformation 
\begin{eqnarray} \label{eq:BdG}
	c_{i\sigma} &=& \sum_{n\kkk}\big(
	u_{n\kkk\sigma}(i)\gamma_{n\kkk\sigma}+
	v_{n\kkk\sigma}^{\ast}(i)\gamma_{n\kkk\bar{\sigma}} \big)e^{-i \kkk \cdot
	\RRR_{i}}\,,
\end{eqnarray}
where the momentum $\kkk$ is contained in the reduced Brillouin zone of the superlattice, and using the fact that
\begin{eqnarray}
	\chi^{+-}_{0}(\mbf{r}_i,\mbf{r}_j,\omega)=\frac{1}{\beta}\int_{0}^{\beta}\mathrm{d}\tau e^{i\omega_n \tau}\chi^{+-}_{0}(\mbf{r}_i,\mbf{r}_j,\tau)
\end{eqnarray}
results in the expression
\begin{eqnarray}
	\chi^{+-}_{0}(\rrr_{i},\rrr_{j},\omega)=\sum_{\kkk \kkk'}f^{0}_{\kkk \kkk'}(i, j)e^{-i(\kkk - \kkk')\cdot(\RRR_{i}-\RRR_{j})}\,, \label{eq:f0}
\end{eqnarray}
where $f^{0}_{\kkk \kkk'}(i,j)$ is the following real space matrix (after analytical continuation, $\omega_n \rightarrow \omega + i\eta$)
\begin{eqnarray}
	f^{0}_{\kkk \kkk'}(i,j) &\!=\!& \frac{-1}{N_s^2}\sum_{nm}\big[ v_{n\kkk\downarrow}(j)u^{\ast}_{n\kkk\uparrow}(i)v^{\ast}_{m\kkk'\downarrow}(i)u_{m\kkk'\uparrow}(j) - u_{n\kkk\uparrow}(j)u^{\ast}_{n\kkk\uparrow}(i)v^{\ast}_{m\kkk'\downarrow}(i)v_{m\kkk'\downarrow}(j) \big]\frac{n_{\text{F}}(E_{n\kkk\uparrow})+n_{\text{F}}(E_{m\kkk'\uparrow})-1}{\omega - E_{n\kkk\uparrow} -E_{m\kkk'\uparrow} + i\eta}
\end{eqnarray}
with $N_s=N_{sx}N_{sy}$ is the number of supercells, and $i$ and $j$ denote sites within one supercell and $\eta$ is an artificial broadening. The real-space Dyson equation yields
\begin{eqnarray}
	\chi_{\text{RPA}}^{+-}(\mbf{r}_i,\mbf{r}_j,\omega) = \chi^{+-}_{0}(\mbf{r}_i,\mbf{r}_j,\omega) + \sum_{\mbf{r}_a}\chi^{+-}_{0}(\mbf{r}_i,\mbf{r}_a,\omega)U\chi_{\text{RPA}}^{+-}(\mbf{r}_a,\mbf{r}_j,\omega)\,,
\end{eqnarray}
and since the dependence of the supercell vector $\mbf{R}$ should be the same for the bare and RPA expressions, this expression can be used to derive an expression for $f^{\text{RPA}}$:
\begin{eqnarray}\label{eq:f_rpa}
	\sum_{\mbf{k}} f^{\text{RPA}}_{\kkk \, \kkk + \ppp} = \Big[\mathds{1} - UN_{s}\sum_{\kkk'} f^{0}_{\kkk' \, \kkk' + \ppp} \Big]^{-1} \sum_{\mbf{k}} f^{0}_{\kkk \, \kkk + \ppp}\,,
\end{eqnarray}
note that this is really just the standard RPA expression in momentum space, but the ``susceptibilities'' now have matrix structure. Application of Eq.~\eqref{eq:f0} for the RPA case results in
\begin{eqnarray}
	\chi^{+-}_{\text{RPA}}(\rrr_{i},\rrr_{j},\omega)=\sum_{\kkk \kkk'}f^{\text{RPA}}_{\kkk \, \kkk'}(i, j)e^{-i(\kkk - \kkk')\cdot(\RRR_{i}-\RRR_{j})}\,.
\end{eqnarray}
To make contact with the cross section as measured by neutrons we take advantage of the fact that neutrons are not a local probe and in the Fourier transform to momentum space we average over sites, as indicated in Eq. (5) in the main text.

\section{Selfconsistent determination of $U$}

When solving for self-consistent striped solutions, an assumption about the periodicity of the stripes has to be made. Hence, for a set of input parameters, the periodicity of the ground state might differ from the input periodicity, implying that the solution given from solving the self-consistent equations is not necessarily the ground state of the system.

In the present case we are interested in stripes with a certain periodicity ($8 \times 2$) and varying magnetization and since we are interested in the dynamics of the spin-waves, ensuring that they remain gapless is critical. The regular iterative selfconsistent approach typically results in saddle points in the free energy due to the enforced periodicity, and we therefore adopt an approach focused on enforcing gapless Goldstone modes. In practice this means that we choose a modulation of the electron densities corresponding to an $8 \times 2$--periodic state and adjust $U$ such that the smallest eigenvalue of
\begin{eqnarray}\label{eq:goldstone_criterion}
	\mathds{1}-UN_{s}\text{Re}\left[ \sum_{\mbf{k}}f^{0}_{\mbf{k}\,\mbf{k+q}}(\mbf{i},\mbf{j},U) \right]\,,
\end{eqnarray}
the real part of the denominator of the RPA susceptibility, is zero. The gapless excitations should appear at $\mbf{q}=(\pi \pm \tfrac{\pi}{4},\pi)$ as the underlying state exhibits $8 \times 2$--periodicity, and $\mu$ is adjusted to ensure that this is the case. Here we included $U$ as a dependent in $f^0$ to remind the reader that $f^0$ also changes as a function of $U$, as is seen from the mean-field Hamiltonian.

This procedure allows us to vary the size of the magnetic moments and the type of superconducting order independently, while remaining in the same parameter regime and simultaneously satisfying Goldstone's theorem. In practice this is carried out by choosing a certain density modulation, i.e. the values of $\langle n_{i\sigma} \rangle$, and solving Eq.~(\ref{eq:goldstone_criterion}) for either no SC order, dSC or PDW order. The chosen density modulations are based on selfconsistent solutions for a given $U$ and superconducting is imposed on top, without accounting for feedback effects. The values obtained for $U$ and $\mu$ by finding the zero eigenvalues of Eq.~(\ref{eq:goldstone_criterion}) will therefore depend slightly on which type of superconducting order (if any) was assumed. Below the exact densities for the three cases (a) $M \sim 0.05$ (Tab.~\ref{tab:case_a}), (b) $M \sim 0.10$ (Tab.~\ref{tab:case_b}) and (c) $M \sim 0.4$ (Tab.~\ref{tab:case_c}) are provided, along with the various values of $U$ and $\mu$ following from this procedure. Note that for cases (a) and (b) a slightly different value of $t'=-0.22$ was used in order for Eq.~\ref{eq:goldstone_criterion} to yield a zero eigenvalue.

\subsection*{Case (a)}
\begin{table}[h]
\centering
\begin{tabular}{|c|c|c|c|c|c|c|c|}
\hline
	0.4070 & 0.4372 & 0.4680 & 0.3948 & 0.4680 & 0.4372 & 0.4070 & 0.4808 \\
\hline
	0.4680 & 0.4372 & 0.4070 & 0.4808 & 0.4070 & 0.4372 & 0.4680 & 0.3948 \\
\hline
\end{tabular}\\
\vspace{5mm}
\begin{tabular}{|c|c|c|c|c|c|c|c|}
\hline
	0.4680 & 0.4372 & 0.4070 & 0.4808 & 0.4070 & 0.4372 & 0.4680 & 0.3948 \\
\hline
	0.4070 & 0.4372 & 0.4680 & 0.3948 & 0.4680 & 0.4372 & 0.4070 & 0.4808 \\
\hline
\end{tabular}\\
\vspace{5mm}
\begin{tabular}{|c|c|c|c|}
\hline
 & non-SC & dSC & PDW \\
\hline
$U$ & 1.9467 & 2.2303 & 1.9074 \\
\hline
$\mu$ & $-0.1$ & 0.131 & $-0.1$ \\
\hline
\end{tabular}
\caption{\label{tab:case_a} (top) $\langle n_{i\uparrow} \rangle$, (middle) $\langle n_{i\downarrow} \rangle$ for the $8 \times 2$ sites in the supercell, and (bottom) parameters resulting in a zero eigenvalue solution of Eq.~\ref{eq:goldstone_criterion} for the various choices of superconducting orders.}
\end{table}
\clearpage
\subsection*{Case (b)}

\begin{table}[h]
\centering
\begin{tabular}{|c|c|c|c|c|c|c|c|}
\hline
	0.3650 & 0.4300 & 0.5100 & 0.3250 & 0.5100 & 0.4300 & 0.3650 & 0.5650 \\
\hline
	0.5100 & 0.4300 & 0.3650 & 0.5650 & 0.3650 & 0.4300 & 0.5100 & 0.3250 \\
\hline
\end{tabular}\\
\vspace{5mm}
\begin{tabular}{|c|c|c|c|c|c|c|c|}
\hline
	0.5100 & 0.4300 & 0.3650 & 0.5650 & 0.3650 & 0.4300 & 0.5100 & 0.3250 \\
\hline	
	0.3650 & 0.4300 & 0.5100 & 0.3250 & 0.5100 & 0.4300 & 0.3650 & 0.5650 \\
\hline
\end{tabular}\\
\vspace{5mm}
\begin{tabular}{|c|c|c|c|}
\hline
 & non-SC & dSC & PDW \\
\hline
$U$ & 2.1519 & 2.3564 & 2.1 \\
\hline
$\mu$ & 0.02 & 0.15 & 0.02 \\
\hline
\end{tabular}
\caption{\label{tab:case_b} (top) $\langle n_{i\uparrow} \rangle$, (middle) $\langle n_{i\downarrow} \rangle$ for the $8 \times 2$ sites in the supercell, and (bottom) parameters resulting in a zero eigenvalue solution of Eq.~\ref{eq:goldstone_criterion} for the various choices of superconducting orders.}
\end{table}

\subsection*{Case (c)}

\begin{table}[h]
\centering
\begin{tabular}{|c|c|c|c|c|c|c|c|}
\hline
	0.2200 & 0.4400 & 0.6600 & 0.1800 & 0.6600 & 0.4400 & 0.2200 & 0.7200 \\
\hline
	0.6600 & 0.4400 & 0.2200 & 0.7200 & 0.2200 & 0.4400 & 0.6600 & 0.1800 \\
\hline
\end{tabular}\\
\vspace{5mm}
\begin{tabular}{|c|c|c|c|c|c|c|c|}
\hline
	0.6600 & 0.4400 & 0.2200 & 0.7200 & 0.2200 & 0.4400 & 0.6600 & 0.1800 \\
\hline	
	0.2200 & 0.4400 & 0.6600 & 0.1800 & 0.6600 & 0.4400 & 0.2200 & 0.7200 \\
\hline
\end{tabular}\\
\vspace{5mm}
\begin{tabular}{|c|c|c|c|}
\hline
 & non-SC & dSC & PDW \\
\hline
$U$ & 3.16 & 3.3097 & 3.2235 \\
\hline
$\mu$ & 0.25 & 0.25 & 0.25 \\
\hline
\end{tabular}
\caption{\label{tab:case_c} (top) $\langle n_{i\uparrow} \rangle$, (middle) $\langle n_{i\downarrow} \rangle$ for the $8 \times 2$ sites in the supercell, and (bottom) parameters resulting in a zero eigenvalue solution of Eq.~\ref{eq:goldstone_criterion} for the various choices of superconducting orders.}
\end{table}

\end{widetext}


\begin{thebibliography}{00}

\bibitem{scalapino12} D J. Scalapino, Rev. Mod. Phys. {\bf 84}, 1383 (2012).
%
\bibitem{berk66} N. F. Berk and J. R. Schrieffer, Phys. Rev. Lett. {\bf 17}, 433 (1966).
%
\bibitem{tranquada95} J. M. Tranquada, B. J. Sternlieb, J. D. Axe, Y. Nakamura, and S. Uchida, Nature (London) {\bf 375}, 561 (1995).
%
\bibitem{klauss00} H.-H. Klauss, W. Wagener, M. Hillberg, W. Kopmann, H. Walf, F. J. Litterst, M. H\"{u}cker, and B. B\"{u}chner, Phys. Rev. Lett. {\bf 85}, 4590 (2000).
%
\bibitem{fujita04} M. Fujita, H. Goka, K. Yamada, J. M. Tranquada, and L. P. Regnault, Phys. Rev. B {\bf 70}, 104517 (2004).
%
\bibitem{fink09} J. Fink, E. Schierle, E. Weschke, J. Geck, D. Hawthorn, V. Soltwisch, H. Wadati, H.-H. Wu, H. A. D\"{u}rr, N. Wizent, B. B\"{u}chner, and G. A. Sawatzky, Phys. Rev. B {\bf 79}, 100502(R) (2009).
%
\bibitem{hucker11} M. H\"{u}cker, M. v. Zimmermann, G. D. Gu, Z. J. Zu, J. S. Wen, G. Xu, H. J. Kang, A. Zheludev, and J. M. Tranquada, Phys. Rev. B {\bf 83}, 104506 (2011).
%
\bibitem{kivelson03} S. A. Kivelson, E. Fradkin, V. Oganesyan, J. M. Tranquada, A. Kapitulnik, and C. Howald, Rev. Mod. Phys. {\bf 75}, 1201 (2003).
%
\bibitem{cupratereview1} J. M. Tranquada, in Handbook of High-Temperature Superconductivity Theory and Experiment, edited by J. R. Schrieffer, (Springer, New York, 2007).
%
\bibitem{cupratereview2} M. Vojta, Adv. Phys. {\bf 58}, 699 (2009).
%
\bibitem{Eschrig2006} M. Eschrig, Adv. Phys. {\bf 55}, 47 (2006).
%
\bibitem{li07} Q. Li, M. H\"{u}cker, G. D. Gu, A. M. Tsvelik, and J. M. Tranquada, Phys. Rev. Lett. {\bf 99}, 067001 (2007).
%
\bibitem{agterberg08} D. F. Agterberg and H. Tsunetsugu, Nat. Phys. {\bf 4}, 639 (2008).
%
\bibitem{barush08} S. Baruch and D. Orgad, Phys. Rev. B {\bf 77}, 174502 (2008).
%
\bibitem{Berg07} E. Berg, E. Fradkin, E.-A. Kim, S. A. Kivelson, V. Oganesyan, J. M. Tranquada, and S. C. Zhang, Phys. Rev. Lett. {\bf 99}, 127003 (2007).
%
\bibitem{berg09} E. Berg, E. Fradkin, and S. A. Kivelson, Phys. Rev. B {\bf 79}, 064515 (2009).
%
\bibitem{Fradkin15} E. Fradkin, S. A. Kivelson, J. M. Tranquada, Rev. Mod. Phys. {\bf 87}, 457 (2015).
%
\bibitem{Himeda02} A. Himeda, T. Kato, M. Ogata, Phys. Rev. Lett. {\bf 88}, 117001 (2002).
%
\bibitem{Raczkowski07} M. Raczkowski, M. Capello, D. Poilblanc, R. Fr{\'e}sard, and A. M. Ole{\'s}, Phys. Rev. B. {\bf 76}, 140505 (2007).
%
\bibitem{Yang09} K.-Y. Yang, W. Q. Chen, T. M. Rice, M. Sigrist, F.-C. Zhang, New J. Phys. {\bf 11}, 055053 (2009). 
%
\bibitem{Corboze14} P. Corboz, T. M. Rice, M. Troyer, Phys. Rev. Lett. {\bf 113}, 046402 (2014).
%
\bibitem{loder10} F. Loder, A.P. Kampf, and T. Kopp, Phys. Rev. B {\bf 81}, 020511(R) (2010).
%
\bibitem{loder11} F. Loder, S. Graser, A. P. Kampf, and T. Kopp, Phys. Rev. Lett. {\bf 107}, 187001 (2011).
%
\bibitem{loder11njp} F. Loder, S. Graser, M. Schmid, A.P. Kampf, and T. Kopp, New J. Phys. {\bf 13}, 113037 (2011).
%
\bibitem{lee14} P. A. Lee, Phys. Rev. X {\bf 4}, 031017 (2014).
%
\bibitem{Wang14} Y. Wang, A. Chubukov, Phys. Rev. B. {\bf 90}, 035149 (2014).
%
\bibitem{Pepin14} C. P\'epin, V. S. de Carvalho, T. Kloss, X. Montiel, Phys. Rev. B. {\bf 90}, 195207 (2014).
%
\bibitem{Freire15} H. Freire, V. S. de Carvalho, C. P\'epin,  Phys. Rev. B. {\bf 92}, 045132 (2015). 
%
\bibitem{Wang15} Y. Wang, D. F. Agterberg, A. Chubukov,  Phys. Rev. Lett. {\bf 114}, 197001 (2015).
%
\bibitem{Wang15_2} Y. Wang, D. F. Agterberg, and A. V. Chubukov, Phys. Rev. B {\bf 91}, 115103 (2015).
%
\bibitem{xu14} Z. Xu, C. Stock, S. Chi, A. I. Kolesnikov, G. Xu, G. Gu, and J. M. Tranquada, Phys. Rev. Lett. {\bf 113}, 177002 (2014).
%
\bibitem{scalapino95} D. J. Scalapino, Phys. Rep. {\bf 250}, 329 (1995).
%

\bibitem{zhu02} J.-X. Zhu, I. Martin, and A. R. Bishop, Phys. Rev. Lett. {\bf 89}, 067003 (2002).
%
\bibitem{Chen} Y. Chen and C. S. Ting, Phys. Rev. Lett. {\bf 92}, 077203 (2004).
%
\bibitem{andersen05} B. M. Andersen and P. Hedeg\aa rd, Phys. Rev. Lett. {\bf 95}, 037002 (2005).
%
\bibitem{Andersen10} B. M. Andersen, S. Graser, and P. J. Hirschfeld, Phys. Rev. Lett. {\bf 105}, 147002 (2010).
%
\bibitem{dallapiazza15} B. Dalla Piazza, M. Mourigal, N. B. Christensen, G. J. Nielsen, P. Tregenna-Piggott, T. G. Perring, M. Enderle, D. F. McMorrow, D. A. Ivanov, and H. M. R{\o}nnow, Nat. Phys. \textbf{11}, 62 (2015).
%
\bibitem{supplementary} Supplementary Material.
%
\bibitem{Andersen06} B. M. Andersen, A. Melikyan, T. S. Nunner, and P. J. Hirschfeld, Phys. Rev. Lett. {\bf 96}, 097004 (2006).
%
\bibitem{footnote} In the two cases in Fig.~\ref{fig:med_M_spin_waves}(a-c) and Fig.~\ref{fig:med_M_spin_waves}(d-f), a slightly different value of $t'=-0.22$ was used to facilitate the satisfaction of Goldstone's theorem.
%
\bibitem{kruger03} F. Kr\"{u}ger and S. Scheidl, Phys. Rev. B {\bf 67}, 134512 (2003).
%
\bibitem{seibold06} G. Seibold and J. Lorenzana, Phys. Rev. B {\bf 73} 144515 (2006).
%
\bibitem{yao06} D. X. Yao, E. W. Carlson, and D. K. Campbell, Phys. Rev. B {\bf 73} 224525 (2006).


\end{thebibliography}
\end{document}